\documentclass[twocolumn,aps,superscriptaddress]{revtex4-2}

\usepackage{graphicx}% Include figure files
\usepackage{bm}% bold math
\usepackage[dvipsnames]{xcolor}
\usepackage[mathlines]{lineno}% Enable numbering of text and display math
\usepackage{braket}
\usepackage{hyperref}
\usepackage{tabularx}
\usepackage[english]{babel}
\usepackage{amsmath}
\usepackage{mathtools}
\usepackage[noend]{algpseudocode}
\usepackage{algorithm}
\usepackage{ulem}

\begin{document}

\title{Real-time adaptive estimation of decoherence timescales for a single qubit}

\author{Muhammad Junaid Arshad}
\affiliation{ 
SUPA, Institute of Photonics and Quantum Sciences, School of Engineering and Physical Sciences, Heriot-Watt University,
Edinburgh EH14 4AS, UK
}

\author{Christiaan Bekker}
\affiliation{ 
SUPA, Institute of Photonics and Quantum Sciences, School of Engineering and Physical Sciences, Heriot-Watt University,
Edinburgh EH14 4AS, UK
}

\author{Ben Haylock}
\affiliation{ 
SUPA, Institute of Photonics and Quantum Sciences, School of Engineering and Physical Sciences, Heriot-Watt University,
Edinburgh EH14 4AS, UK
}

\author{Krzysztof Skrzypczak}
\affiliation{ 
SUPA, Institute of Photonics and Quantum Sciences, School of Engineering and Physical Sciences, Heriot-Watt University,
Edinburgh EH14 4AS, UK
}

\author{Daniel White}
\affiliation{ 
SUPA, Institute of Photonics and Quantum Sciences, School of Engineering and Physical Sciences, Heriot-Watt University,
Edinburgh EH14 4AS, UK
}

\author{Benjamin Griffiths}
\affiliation{Department of Materials, University of Oxford, Parks Road, Oxford OX1 3PH, UK}

\author {Joe Gore}
\affiliation{Department of Physics, University of Warwick, Coventry CV4 7AL, UK}

\author {Gavin W. Morley}
\affiliation{Department of Physics, University of Warwick, Coventry CV4 7AL, UK}

\author {Patrick Salter}
\affiliation{Department of Engineering Science, University of Oxford, Parks Road, Oxford OX1 3PJ, UK}

\author{Jason Smith}
\affiliation{Department of Materials, University of Oxford, Parks Road, Oxford OX1 3PH, UK}

\author{Inbar Zohar}
\affiliation{Department of Chemical and Biological Physics
Weizmann Institute of Science, Rehovot 7610001, Israel}

\author{Amit Finkler}
\affiliation{Department of Chemical and Biological Physics
Weizmann Institute of Science, Rehovot 7610001, Israel}

\author{Yoann Altmann}
\affiliation{ 
Institute of Signals, Sensors and Systems, School of Engineering and Physical Sciences, Heriot-Watt University,
Edinburgh EH14 4AS, UK
}
\author{Erik M. Gauger}
\affiliation{ 
SUPA, Institute of Photonics and Quantum Sciences, School of Engineering and Physical Sciences, Heriot-Watt University,
Edinburgh EH14 4AS, UK
}

\author{Cristian Bonato}
\email{c.bonato@hw.ac.uk}
\affiliation{ 
SUPA, Institute of Photonics and Quantum Sciences, School of Engineering and Physical Sciences, Heriot-Watt University,
Edinburgh EH14 4AS, UK
}

\begin{abstract}

Characterising the time over which quantum coherence survives is critical for any implementation of quantum bits, memories and sensors. The usual method for determining a quantum system's decoherence rate involves a suite of experiments probing the entire expected range of this parameter, and extracting the resulting estimation in post-processing. Here we present an adaptive multi-parameter Bayesian approach, based on a simple analytical update rule, to estimate the key decoherence timescales ($T_1$, $T_2^*$ and $T_2$) and the corresponding decay exponent of a quantum system in real time, using information gained in preceding experiments. This approach reduces the time required to reach a given uncertainty by a factor up to an order of magnitude, depending on the specific experiment, compared to the standard protocol of curve fitting. A further speed-up of a factor $\sim 2$ can be realised by performing our optimisation with respect to sensitivity as opposed to variance.

\end{abstract}

%Use showkeys class option if keyword
\maketitle

\section{\label{sec:level1} Introduction}

Decoherence, resulting from the interaction of a quantum system with its environment, is a key performance indicator for qubits in quantum technologies \cite{suter2016colloquium} including quantum communication, computation and sensing.
Decoherence timescales determine the storage time for quantum memories and quantum repeaters, a crucial metric for quantum communication networks \cite{wehner2018quantum,pompili2021realization,bradley2019ten,zhong2015optically}. Rapid benchmarking of decoherence timescales in platforms such as superconducting qubits \cite{arute2019quantum,wu2021strong} or silicon spin qubits \cite{crippa2019gate,vinet2021path}, is a critical validation and quality assurance step for the development of large-scale quantum computing architectures, and has the potential to improve error correction protocols efficiently close to fault-tolerance thresholds. In quantum sensing, the role of decoherence is two-fold. On one side, decoherence sets the ultimate performance limit of the sensors \cite{degen2017quantum}. On the other hand, decoherence itself can be the quantity measured by a quantum sensor, as it provides information about the environment. An example of this is relaxometry, where the rate at which a polarised quantum sensor reaches the thermal equilibrium configuration gives information about different physical processes in the environment \cite{schmid_lorch_2015, finco2021imaging, nie2021quantum, zhang2021toward}.

Decoherence rates can be measured by preparing the system into a known quantum state and probing it at varying time delays to determine the probability of decay from its initial state. The standard protocol for decoherence estimation involves a series of measurements with time delays set over a pre-determined range, reflecting the expected value of the decoherence rate, and fitting of the result to a decay function. As the range of time delays is determined in advance, some of the measurements will provide little information on the decoherence of the system, since the time delay is either much shorter than the true decoherence rate, resulting in no decay, or much longer, resulting in complete decay.

Here we introduce an real-time adaptive protocol to measure decoherence timescales $T_1$, $T_2^*$ and $T_2$ for a single qubit \cite{zhang2018improving}, respectively corresponding to relaxation, dephasing, and echo decay time \cite{suter2016colloquium}, together with the coherence decay exponent $\beta$. While the proposed algorithms are very general and can be applied to any quantum architecture, our experiments are implemented on a single spin qubit associated with a nitrogen-vacancy (NV) centre in diamond.

Adaptive techniques have been shown to be central to progress across a broad range of quantum technologies \cite{gebhart2023learning}. Early work in this field involved the implementation of adaptive quantum phase estimation algorithms on photonic systems \cite{higgins2007entanglement}, later extended to frequency estimation with applications to (static) DC magnetometry with single electron spins \cite{said2011nanoscale, bonato2016optimized, Zohar2023}. Alternative adaptive protocols for the estimation of static magnetic fields are based on sequential Bayesian experiment design \cite{mcmichael2021sequential, mcmichael_simplified} and ad-hoc heuristics \cite{santagati2019magnetic}, later applied to the characterisation of a single nuclear spin \cite{joas2021online}. Real-time adaptation of experimental settings has also been shown to be advantageous when measuring spin relaxation \cite{caouette2022robust} or tracking the magnetic resonance of a single electron spin in real-time \cite{ambal2019differential,bonato2017adaptive1, craigie2021resource,benestad2023efficient}. 
Furthermore, adaptive techniques have been investigated to enhance photonic quantum sensors \cite{valeri2023experimental,lumino2018experimental,cimini2023deep}, as a control tool for quantum state preparation \cite{blok2014manipulating} and to extend quantum coherence of a qubit by manipulating the environment \cite{vijay2012stabilizing,scerri2020extending,shulman2014suppressing}.

Despite these pioneering experiments, several important methodological questions still remain open. A priority concern is that adaptive protocols introduce an overhead, given by the time required to compute settings on the fly for the next iteration. It is crucial to minimise this computation time, since it can slow the protocol down to the point that the overhead can reverse the gain in measurement speed compared to a simple parameter sweep. This has not been considered in many cases, in particular where algorithms were investigated through computer simulations \cite{bonato2017adaptive1, dinani2019, mcmichael2021sequential, mcmichael_simplified} or as off-line processing of pre-existing experimental data \cite{santagati2019magnetic}. While the optimisation of complex utility functions can possibly deliver the best  theoretical results, this could be practically less advantageous than near-optimal approaches with very fast update rules in minimising total measurement duration. A second issue is related to the fact that, for multi-parameter Hamiltonian estimation, standard approaches such as the maximisation of Fisher information can fail, as the Fisher information matrix becomes singular when controlling the evolution time \cite{granade_hamiltonian_learning_NJP}. This has stimulated researchers to find ad-hoc heuristics, for example, the particle guess heuristic \cite{granade_hamiltonian_learning_NJP, santagati2019magnetic, joas2021online} for the estimation of Hamiltonian terms; these heuristics, however, do not necessarily work beyond Hamiltonian estimation. A third question is related to what quantity should be optimised. Previous work has targeted the minimisation of the variance of the probability distribution for the quantity of interest \cite{cappellaro_PRA_2012, joas2021online}. While this is clear when all measurements feature the same duration, the answer is less straightforward when adapting the probing time. If two measurements with different probing times result in a similar variance, the protocol should prefer the shorter one, minimising the overall sensing time.

Here we address these open questions, presenting theoretical and experimental data about the adaptive estimation of decoherence for a single qubit, using NV centres as a case study. Compared to other recent investigations of adaptive protocols \cite{santagati2019magnetic, joas2021online, caouette2022robust}, our experiments utilize a very simple analytical update rule based on the concept of Fisher information and the Cram\'er-Rao bound. By exploiting state-of-the-art fast electronics, we experimentally perform the real-time processing in than $50 \mu$s, an order of magnitude shorter than previous real-time experiments \cite{joas2021online}, negligible compared to the duration of each measurement. Such a short timescale makes our approach useful for qubits where fast single-shot readout is available such as trapped ions \cite{ion_SSRO}, superconducting qubits \cite{mallet2009single} and several types of spin qubits \cite{hanson_SSRO_QD, robledo2011high, anderson_SSRO_SiC, thompson_parallel_SSRO}, and could be further shortened in future work by implementing the protocols on field-programmable gate array (FPGA) hardware.  

In the case of multi-parameter estimation, previous work on Hamiltonian estimation had pointed out that the Cram\'er-Rao bound cannot be used in the optimisation as the Fisher information matrix is singular and cannot be inverted \cite{granade_hamiltonian_learning_NJP}. Here we address this issue by utilising multiple probing times, showing that the Fisher information matrix can be inverted and that the corresponding adaptive scheme provides better performance than non-adaptive approaches. Finally, we discuss what quantity needs to be targeted to achieve the best sensor performance, experimentally demonstrating the superiority of optimizing sensitivity, defined as variance multiplied by time, over optimizing variance. As a figure of merit, sensitivity encourages faster measurements.

Our work tackles these general questions using the characterisation of decoherence as a test case. While adaptive approaches have been investigated in the case of phase and frequency estimation \cite{higgins2007entanglement, said2011nanoscale, bonato2016optimized, santagati2019magnetic, mcmichael2021sequential, mcmichael_simplified, joas2021online, Zohar2023}, also in relation to Hamiltonian learning \cite{santagati2019magnetic}, the case of decoherence is much less explored, with only one work targeting the estimation of the relaxation timescale $T_1$ \cite{caouette2022robust}. Here we provide the first complete characterisation of the three decoherence timescales typically used in experiments ($T_1$, $T_2^*$ and $T_2$), together with the decoherence decay exponent $\beta$.

\begin{figure*}
\includegraphics[height=0.5\textheight, width=1\textwidth ]{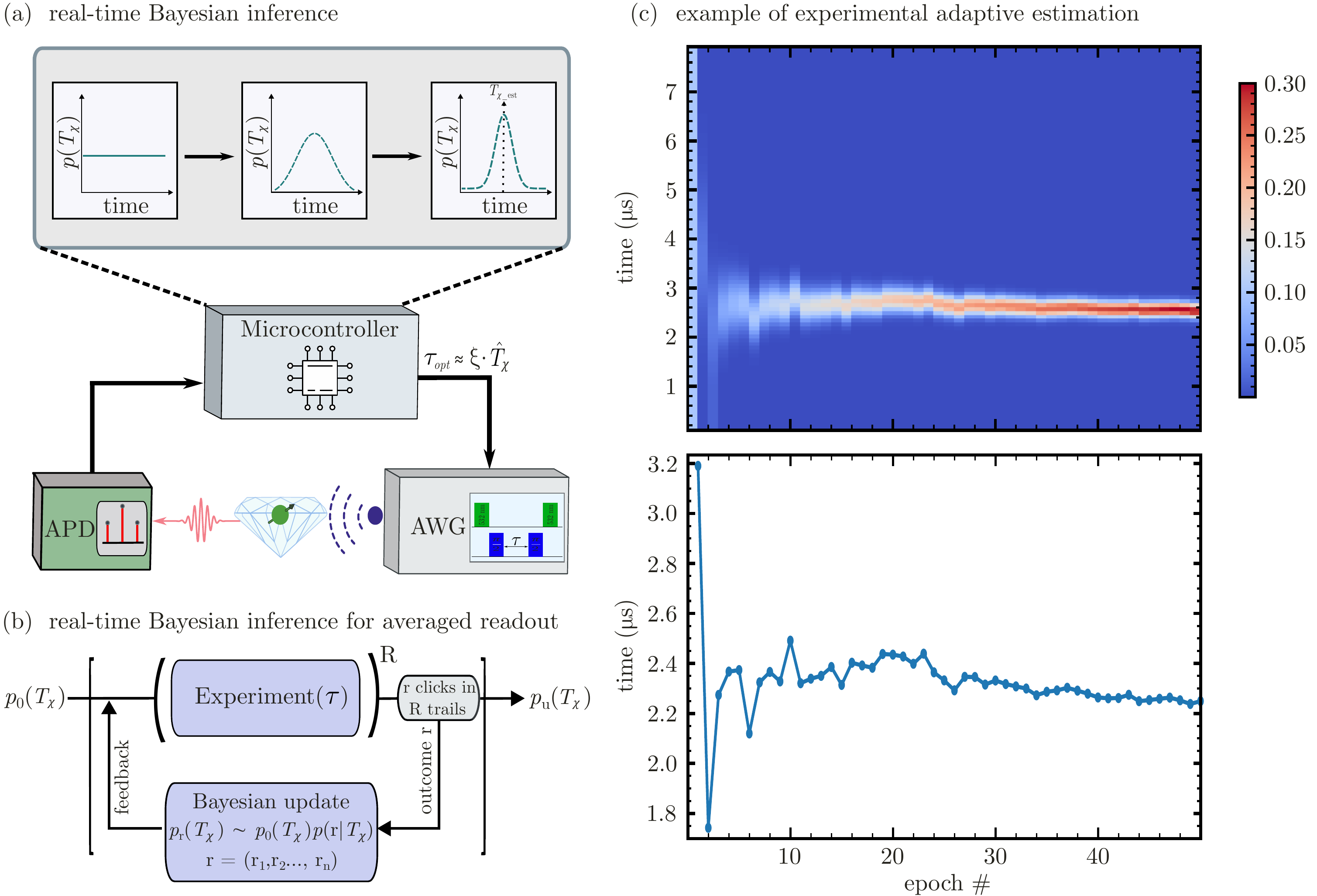}
\caption{\label{fig:1a}\textbf{Real-time adaptive feedback.} (a) Schematic of the real-time adaptive protocol demonstrated in this work, using the electronic spin associated with a nitrogen-vacancy (NV) centre in diamond. An Arbitrary Waveform Generator (AWG) is used to generate pulses for manipulation of the spin qubit. The spin state is then optically measured and the detected photon count rate is used by the microcontroller to estimate the value of the decoherence timescale $T_\chi$ (and the decay exponent $\beta$) using Bayesian inference. The microcontroller also computes the optimal probing time $\tau$ for the next measurement, passing the value to the AWG, which builds the next estimation sequence accordingly. (b) In our experiment, a single interrogation of the qubit does not provide sufficient information to discriminate qubit state. Hence $R$ measurements are performed, and the resulting number $r$ of detected photons are used to update $p (T_{\chi})$ through Bayes' rule and to compute $\tau$ for the next experiment. (c) Example of an experimental adaptive estimation sequence, with $p (T_2^*)$ shown for each measurement epoch (here the decay exponent is known a-priori as $\beta=2$). The probability $p (T_2^*$), initially uniform, converges towards a narrow peak as more measurement outcomes are accumulated. The experimental adaptively-optimised values for $\tau$ are shown on the bottom plot.}
\end{figure*}

\section{Theory}
\label{sec:theory}
Decoherence and relaxation are processes induced by the interaction of a qubit with its environment, leading to random transitions between states or random phase accumulation during the evolution of the qubit. These processes are typically estimated by preparing a quantum state and tracking the probability of still measuring the initial state over time, which can be captured by a functional form \cite{degen2017quantum}

\begin{equation}
    p(t) \propto \frac{1}{2} \left( 1 - e^{-\chi(t)} \right) \quad .
\end{equation}

Although the noise processes induced by interaction with the environment can be complex, $\chi(t)$ can often be approximated by a simple power law:
\begin{equation}
    \chi (t) \propto \left( \frac{t}{T_{\chi}}\right)^{\beta} \quad ,
\end{equation}
where $T_{\chi}$ and $\beta$ depend on the specific noise process \cite{suter2016colloquium}. For white noise, the decay is exponential with $\beta = 1$. For a generic $1/f^q$ decay, relevant for example for superconducting qubits, with a noise spectral density as $\propto \omega^q $, $\chi (t)$ scales as $\chi(t) \propto \left( t/T_{\chi} \right)^{1+q}$ \cite{cywinski2008enhance}. 

In the case of a single electronic spin dipolarly coupled to a diluted bath of nuclear spins, the decay exponents have been thoroughly investigated, with analytical solutions available for different parameter regimes \cite{hall2014analytic}. If the intra-bath coupling can be neglected, the free induction decay of a single spin is approximately Gaussian ($\beta = 2$) \cite{dobrovitski2008decoherence,hanson2008coherent}. The Hahn echo decay exponent $T_2$ can vary, typically between $\beta \sim 1.5-4$ depending on the specific bath parameters and applied static magnetic field \cite{hall2014analytic,bauch2020decoherence}.

In Sec. \ref{sec:single_par_estimation_theory}, we assume the decay exponent $\beta$ to be known, and we only focus on estimating the decay time $T_{\chi}$. This is a practically-relevant situation in cases where the nature of the bath is well-understood and the decay exponent $\beta$ is known, at least approximately, \textit{a priori}. We then extend our analysis to the simultaneous estimation of $T_{\chi}$ and $\beta$ in Sec. \ref{sec:multuiparameter_theory}.

Fig. \ref{fig:1a}(a) sketches the operation of the real-time adaptive sensing system developed in our study. We have utilised the electronic spin associated with a nitrogen-vacancy (NV) center in diamond as the qubit, which is initialised and readout by optical pulses. The qubit state is manipulated by microwave (MW) pulses, created in real-time by an Arbitrary Waveform Generator (AWG) based on an external digital input. After the application of a pulse sequence, the qubit is optically readout, with the spin state information enconded in the number photoluminescence photon counts during optical illumination. The core of our adaptive system is a real-time microcontroller, which uses the detected photon count rate to estimate the values of the decoherence timescale $(T_\chi)$ and the decay exponent $(\beta)$ via Bayesian inference. As shown in the inset, the probability distribution starts out as uniformly flat, but begins to converge around the true value after a few iterations. Based on the estimated value in the current iteration, the microcontroller computes the optimal probing time $(\tau)$ for the subsequent measurement and communicates this value to the AWG, which then constructs the next estimation sequence accordingly. This cycle repeats for several iterations until a desired level of error in the estimation of the target quantity is reached. Fig. \ref{fig:1a}(b) shows the flow of the experimental estimation sequence. For our experiments, a single measurement of the qubit lacks the information required for discriminating its state effectively. Therefore, we conduct R measurements, to obtain r detected photons, enough to discriminate the spin state. Such counts are then utilised to update the probability distribution $p(T_\chi)$ using Bayes' rule. After the Bayesian update, updated probability distribution is used to compute the optimal settings and provide feedback for the subsequent measurements. Fig. \ref{fig:1a}(c) shows an example of experimental estimation of  $T^*_2$, performed by an adaptive Ramsey experiment, plotted as the evolution of $p(T_2^*)$ for increasing estimation epochs. In the beginning, $p (T_2^*)$ is a uniform distribution in the range 0-8 $\mu s$, which then converges to a singly-peaked distribution after more and more measurement outcomes are processed. In the case of an NV centre in a high-purity diamond, the decay is expected to be Gaussian ($\beta=2$) \cite{hanson2008coherent}. As described later in Sec. \ref{sec:single_par_estimation_theory}, the optimal adaptive rule for this case is to choose the probing time as $\tau_{opt} \sim 0.89 \cdot \hat{T}_2^*$ (see Eq. \ref{eq:update_rule}, where $\hat{T}_2^*$ is the current estimate of $T_2^*$ computed from the probability distribution $p (T_2^*)$. The chosen values for $\tau$ are shown on the bottom plot, illustrating how they converge very fast to the optimal value $\tau_{opt} \sim 0.89 \cdot \left(T_2^*\right)_{true} \sim 2.23~ ~\mu$s.

\begin{figure*}
\includegraphics[width=1\textwidth]{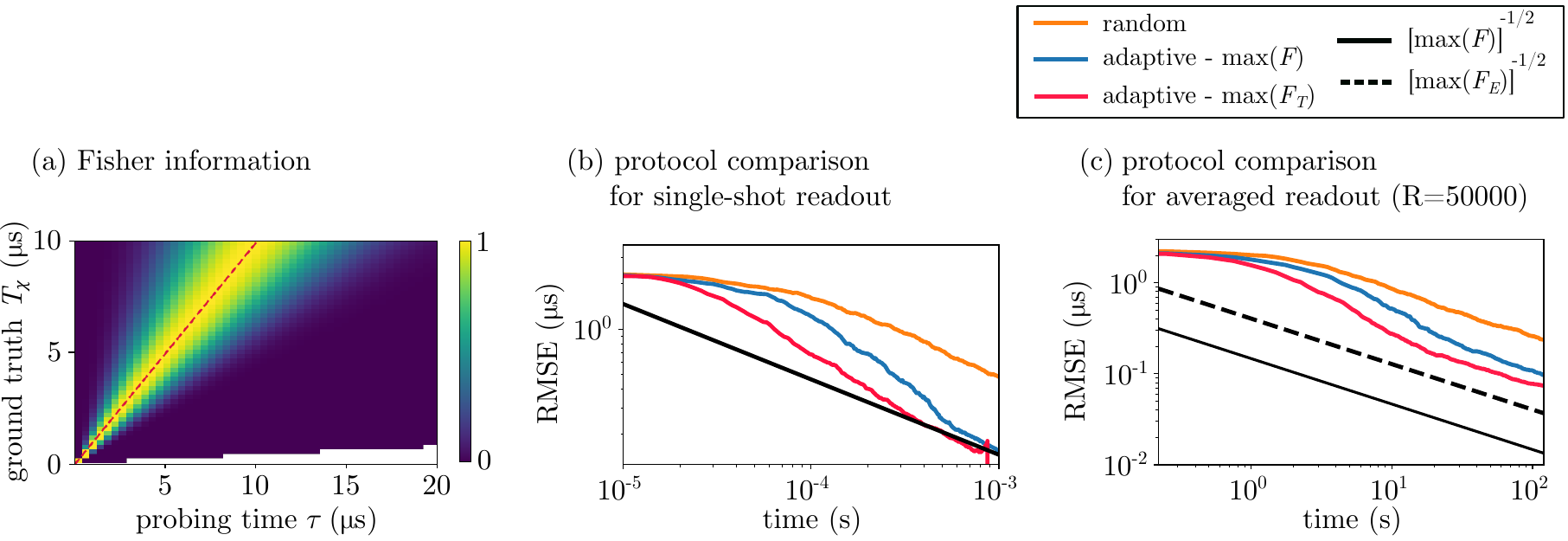}
\caption{\textbf{Numerical simulations for $T_2^*$ estimation.} (a) Fisher information $F$ as a function of the ground truth value for $T_{\chi}$ = $T_2^*$ and the probing time $\tau$, from Eq. (\ref{eq:FI}). The plot shows that the maximum of $F(\tau)$ has a linear dependence on $T_2^*$, following Eq. (\ref{eq:FI}) (shown as the red dashed line). (b) Comparison between different strategies to learn $T_2^*$, assuming that single shot qubit readout is available: random choice of $\tau$ (orange), choice of $\tau$ obtained maximizing the Fisher information $F$ in Eq. (\ref{eq:FI}) (blue) and choice of $\tau$ obtained maximizing the rescaled Fisher information $F_T$ in Eq. (\ref{eq:F_T}) (red). The x-axis shows the total probing time (cumulative over epochs), while the y-axis shows the uncertainty, defined as the root mean squared error (RMSE) from the ground truth value. The solid black line corresponds to the theoretical limit set by the Cram\'er-Rao lower bound (CRLB) for the Fisher information in Eq. (\ref{eq:FI}). Adaptive protocols outperform the non-adaptive (random) protocol. (c) Simulations for the same protocols as in (b), for the case where single shot readout is not available (using experimental values $p_{cl}(\ket{0}) = 0.0187$ and $p_{cl}(\ket{1}) = 0.0148$, and $R = 50000$).  The solid black line corresponds to the limit set by the CRLB for the Fisher information in Eq. (\ref{eq:FI}). The dashed black line shows the CRLB limit for the Fisher information $F_E$ in Eq. (\ref{eq:F_E}).}
\label{fig:T2star_sims_SSRO}
\end{figure*}
.
\subsection{Adaptive Bayesian estimation}
\label{sec:single_par_estimation_theory}
We utilize Bayesian inference, exploiting Bayes' theorem to update knowledge about the decoherence time $T_{\chi}$ and decay exponent $\beta$ in the light of a set of new measurement outcomes denoted by $\Vec{m}$ = \{$m_1$,$m_2$, \ldots\}. Thanks to its flexibility in accounting for experimental imperfections and for integrating real-time adaptation of the experimental setting while remaining easy to interpret mathematically, the Bayesian framework \cite{gebhart2023learning,pezze_entropy_2018} has been widely applied in quantum technology, from sensing \cite{bonato2016optimized,santagati2019magnetic,joas2021online,caouette2022robust}, to the tuning of quantum circuits \cite{lennon2019efficiently,teske2019machine}, and model learning \cite{gentile2021learning}. In this section, we will restrict the discussion to the characterisation of the decoherence time $T_{\chi}$; the extension to a multi-parameter case, with the simultaneous estimation of $T_\chi$ and $\beta$, will be presented in Sec. \ref{sec:multuiparameter_theory}.

For each binary measurement outcome $m_n$ ($m_n=0, 1$), the probability distribution of $T_{\chi}$, which represents our knowledge about $T_{\chi}$, is updated as
\begin{equation}
\label{eq:probability}
    P(T_\chi|m_{1:n}) \propto P(m_n|T_\chi)P(T_\chi|m_{1:(n-1)}),
\end{equation}
where $m_{1:n}=\{m_1,\ldots,m_n\}$.
Here, $P(T_\chi|m_{1:(n-1)})$ is  the posterior probability after $(n-1)$-th update and proceeds to serve as prior distribution at the $n$-th iteration, and $P(m_n|T_\chi)$ is the likelihood function 

\begin{equation}
\label{eq:likelihood}
    P(m|T_{\chi}) = \frac{1 + e^{im\pi} e^{-(\tau / T_{\chi})^{\beta}}}{2} \qquad .
\end{equation}

Note that this likelihood depends on $\tau$ (which we will adjust later) but this dependency is omitted in $P(m|T_{\chi})$ to simplify the notation.
Our approach to adaptive estimation is to derive a simple expression for the optimal parameter settings, that can be computed in real-time by an analytical formula without adding much extra computation time to the sensing process.

A conventional approach to updating $\tau$ adaptively would be to use the information gain as criterion \cite{lennon2019efficiently,lindley1956measure,mcmichael2021sequential}. However, this involves integrals (with respect to $T_{\chi}$) requiring numerical evaluation and an associated significant computational overhead. Here, we instead employ an approximation of the Bayesian information matrix (BIM) (a $1\times 1$ matrix, in the case of a single parameter) \cite{van2007bayesian} which links to the classical Fisher information \cite{steven1993fundamentals}. While computing the BIM also requires the computation of an integral (more precisely an expectation) with respect to $T_{\chi}$, this can now be easily approximated as explained in \hyperref[app:derivations]{Appendix A}.

The Cram\'er-Rao lower bound (CRLB) of $T_{\chi}$ represents the minimum reachable variance for any (unbiased) estimator of $T_{\chi}$, and is inversely proportional to the Fisher information $F$. Thus, maximizing $F$ with respect to the control parameter $\tau$ is expected to improve our estimate of $T_{\chi}$. 

As shown in \hyperref[app:derivations]{Appendix A}, we examine a Bayesian form of the CRLB, computing the corresponding Fisher information $F_{\mathcal{B}}$ can be computed as:

\begin{eqnarray}
    F_{\mathcal{B}}(\tau) \approx \frac{\beta^2 (\tau/\hat{T}_{\chi})^{2\beta}}{\hat{T}_{\chi}^2 \left[ e^{\left(2\tau/\hat{T}_{\chi}\right)^{\beta}}-1\right]}
\end{eqnarray}
where $\hat{T}_{\chi}$ is a point estimate of $T_{\chi}$ before each measurement.

While we are unable to maximize $\hat{F}_{\mathcal{B}}(\tau)$ analytically, approximate solutions exist. We found that the heuristic 
\begin {equation}
\label {eq:update_rule}
 \tau_{opt} \approx \xi \cdot \hat{T}_{\chi}
\end{equation}
leads to satisfactory results, where $\xi$ is a parameter that depends on $\beta$. Some numerically-computed values for $\xi$ are listed in Table I, for some common values of $\beta$. 

The Fisher information $F$ as a function of the ground truth value for $T_{\chi} = T_2^*$ and the probing time $\tau$, from Eq. \eqref{eq:FI} is plotted in Fig. \ref{fig:T2star_sims_SSRO}(a). $F(\tau, T_2^*)$ is normalised by its maximum with respect to $\tau$, for each value of $T_2^*$. The plot shows clearly that the maximum of $F(\tau)$ has a linear dependence on $T_2^*$, following Eq. (\ref{eq:FI}) (shown as the red dashed line).

\begin{table}[h] 
\centering
\label{table:1}
\begin{tabular}{m{0.075\textwidth}m{0.075\textwidth}m{0.075\textwidth}m{0.075\textwidth}m{0.075\textwidth}} 
\hline\hline %inserts double horizontal lines
\textbf{$\beta$} & 1 & 3/2&2&3 \\ [0.5ex] % inserts table
%heading
\hline % inserts single horizontal line
\textbf{$\xi$ ($F$)} &  \mbox{0.79}&\mbox{0.86}&\mbox{0.89}&\mbox{0.92} \\
\textbf{$\xi$ ($F_T$)} & \mbox{-}&\mbox{0.30}&\mbox{0.66}&\mbox{0.85} \\ [1ex] % [1ex] adds vertical space
\hline %inserts single line
\end{tabular}
\caption{ \textbf{Choice of optimal probing time.} The optimal probing time $\tau_{opt}$ is chosen according to Eq. \eqref{eq:update_rule}, with factor $\xi$. Numerically computed values of $\xi$ for different decay exponents $\beta$ are shown in the table, computed by maximizing either the classical Fisher information $F$ [see Eq. \eqref{eq:FI}] or the rescaled Fisher information $F_T$ [see Eq. \eqref{eq:F_T}]. No maximum of $F_T$ can be found for exponential decay, when $\beta = 1$.}
\label{FI_table}
\end{table}

\subsection{\label{sec:multuiparameter_theory} Multi-parameter estimation}

In many practical situations, it is important to learn both the decoherence timescale, $T_{\chi}$, and the noise exponent, $\beta$, as the latter provides useful information about the nature of the qubit environment.

Directly extending the approach for a single parameter estimate is unfruitful as the determinant of the BIM is zero, as reported previously \cite{granade_hamiltonian_learning_NJP}. The intuitive reason for this is that the determination of two parameters from measurements in one single setting creates correlations between the two parameters, resulting in a singular Fisher information matrix. 

We address this issue by using two consecutive measurement results, m and n, taken at times $\tau_0$ and $\tau_1$. In this case, a non-zero determinant can be found by maximising the BIM on $p(m|T_{\chi},\beta)\times p(n|T_{\chi}, \beta)$. Assuming binary measurement outcomes, the determinant of this matrix is given by:
\begin{widetext}
\begin{equation}
    \det \hat{F}_{\mathcal{B}} = \dfrac{\beta^2\left(\dfrac{\tau_0}{T_{\chi}}\right)^{2\beta}\left(\dfrac{\tau_1}{T_{\chi}}\right)^{2\beta}\left(\log^2\left(\dfrac{\tau_0}{T_{\chi}}\right) - 2\log\left(\dfrac{\tau_0}{T_{\chi}}\right)\log\left(\dfrac{\tau_1}{T_{\chi}}\right) + \log^2\left(\dfrac{\tau_1}{T_{\chi}}\right)\right)}{T_{\chi}^2\left(-\exp\left[2\left(\dfrac{\tau_0}{T_{\chi}}\right)^\beta\right] - \exp{\left[2\left(\dfrac{\tau_1}{T_{\chi}}\right)^\beta\right]} + \exp{\left[2\left(\dfrac{\tau_0}{T_{\chi}}\right)^\beta + 2\left(\dfrac{\tau_1}{T_{\chi}}\right)^\beta\right]} + 1\right)}
    \label{eq:detBIM2d4o}
\end{equation}
\end{widetext}

We did not find a way of calculating the maxima of this function analytically. Instead, here it is approximated numerically to determine the update heuristic. Starting by numerically estimating the value of $\tau_1/T_{\chi}$, which maximises Eq. \eqref{eq:detBIM2d4o} across a range of values of $\tau_0/T_{\chi}$ and $\beta$, we can fit a piece-wise linear approximation for the best choice of $\tau_1$ that works well for $1<\beta<5$. This approximation is given by:
\begin{equation}
    \tau_{1_{opt}} = \begin{dcases}
    0.313\tau_0 +1.04\hat{T}_{\chi}, & \text{if} ~~~\tau_0 < 0.83\hat{T}_{\chi}\\
    0.7\hat{T}_{\chi}, & \text{if} ~~~ 0.83\hat{T}_{\chi} < \tau_0  < 0.96\hat{T}_{\chi} \\
    0.109\tau_0 + 0.55\hat{T}_{\chi}, & \text{if} ~~~ 0.96\hat{T}_{\chi} < \tau_0
    \end{dcases}
    \label{eq:2paramapprox}
\end{equation}

\subsection{\label{sec:opt_sens} Optimizing sensitivity}
The approach we introduced in the previous section aims at maximizing the BIM as a proxy for minimizing the uncertainty of $T_\chi$. However, this criterion does not account for the overall measurement duration. For example, if two measurements (obtained from two different values of $\tau$) lead to similar BIMs, the shorter one of the two should be more favourable, as our goal is to reach the smallest possible uncertainty in the shortest time.
We therefore also consider an alternative approach which, instead of targeting the minimisation of the mean-squared error (MSE), is designed to improve the sensitivity, defined as $\eta^2 = MSE \cdot \tau$ \cite{dinani2019}. This leads to a new criterion that is the BIM rescaled by the probing time:

\begin{equation}
\label{eq:F_T}
    F_T = \dfrac{\hat{F}_{\mathcal{B}}(\tau)}{\tau}=\frac{{\beta}^2 (\tau/\hat{T}_{\chi})^{2\beta}}{\tau \hat{T}_{\chi}^2  \left[e^{2(\tau/\hat{T}_{\chi})^{\beta}}-1\right]} \qquad.
\end{equation}

Again, numerically solving for $(\tau/\hat{T}_{\chi})$, we can find the optimal value $\tau_{opt} = \xi (F_T) \cdot \hat{T}_{\chi}$. The bottom row in Table \ref{FI_table} shows different values for the multiplication factor $\xi$ for different $\beta$ values. We are unable to find a maximum for the re-scaled Fisher information when $\beta = 1$.

\subsection{Theoretical limits}

The fundamental performance limit for decoherence timescale learning is given by the CRLB for the Fisher information in Eq. (\ref{eq:FI}). 
Simulations for the performance of our proposed algorithm for perfect single-shot readout are presented in Fig. \ref{fig:T2star_sims_SSRO}(b), comparing three strategies to learn $T_2^*$: random choice of $\tau$ (orange curve), choice of $\tau$ obtained maximizing the Fisher information $F$ in Eq. (\ref{eq:FI}) (i.e.~minimizing the lower bound on the variance, blue curve) and choice of $\tau$ obtained maximizing the rescaled Fisher information $F_T$ in Eq. (\ref{eq:F_T}) (i.e.~minimizing the `sensitivity', red curve). The solid black line corresponds to the theoretical limit set by the Cram\'er-Rao lower bound (CRLB) for the Fisher information expression in Eq. (\ref{eq:FI}). The curves represent the performance averaged over 500 repetitions of each protocol. The simulations show that both adaptive approaches outperform the non-adaptive (random $\tau$) protocol (orange solid line), with the adaptive protocol maximizing $F_T$ (red line) performing better than the maximisation of $F$ (blue line). Numerical simulations in Fig. \ref{fig:T2star_sims_SSRO}(b) confirm that the estimator is asymptotically unbiased, as the root-mean-square-error (RMSE) becomes smaller and smaller with increasing sensing time, approaching the fundamental estimation limit set by the Cramer-Rao bound.

To disentangle the effect of sub-optimal readout from the performance of the proposed estimation algorithm, we can compute the best performance achievable for a given readout strategy. This can be computed from the Cram\'er-Rao bound, considering the classical Fisher information $F_E$ for the experimental likelihood given by Eq. (\ref{eq:p_D}):  
\begin{widetext}
\begin{equation}
\label{eq:F_E}
    F_E (\tau) = - \frac{\alpha V^2 \beta^2  \left(\frac{\tau}{T_{\chi}}\right)^{2 \beta}}{T_{\chi}^{2} \left(\alpha V^2 + 2\alpha V e^{\left(\frac{\tau}{T_{\chi}}\right)^{\beta}} - V e^{\left(\frac{\tau}{T_{\chi}}\right)^{\beta}} + \alpha e^{2 \left(\frac{\tau}{T_{\chi}}\right)^{\beta}} - e^{2 \left(\frac{\tau}{T_{\chi}}\right)^{\beta}}\right)} \,.
\end{equation}
\end{widetext}
Once again, the ultimate lower bound on the variance obtainable in the limit of many identical repetitions (epochs) is proportional to the inverse $F_E$. To find the optimal probing time, given experimental values for $\alpha$ and $V$, we, therefore, minimize $1/F_E$, displayed as the solid black line in Fig. \ref{fig:T2star_sims_SSRO}(b) and \ref{fig:T2star_sims_SSRO}(c).

\section{Experimental implementation}
\setcounter{subsection}{0}
\subsection{Setup and estimation algorithm}
\label{subsec:Algo}
We demonstrate the proposed protocols using the single electron spin associated with a nitrogen-vacancy (NV) centre in diamond \cite{schirhagl2014nitrogen,barry2020sensitivity}, created by a laser writing method \cite{chen2019laser, stephen2019deep}. The NV electron spin is polarised and measured optically, even at room temperature; it can be further controlled by microwave pulses.

Whilst optical measurement at room temperature is available, it is not possible to obtain binary single shot read-out, unlike at cryogenic temperatures \cite{robledo2011high}. The absence of a single shot readout can be circumvented by repeating each measurement sequence $R$ times. At the $n$-th iteration, the number $r_n$ of detected photons is then used to update the distribution of $T_{\chi}$ as %$p(T_{\chi})$ as

\begin{equation}
    p (T_{\chi}| r_{(1:n)}, R) \propto p (r_n | T_{\chi}, R) p (T_{\chi}| r_{1:(n-1)}, R).
\end{equation}
The likelihood is a binomial distribution (as a result of $R$ independent and identical Bernoulli experiments). The probability of detection of a photon click in a single repetition is given by
\begin{equation}
\label{eq:p_D}
    p_D (\tau, T_{\chi}, \beta) = \alpha \left(1 + V e^{-(\tau/T_{\chi})^\beta} \right) \,.
\end{equation}
This probability depends on $V$ and $\alpha$, which can be computed from the experimental probabilities to detect a click when the qubit is in $\ket{0}$ ($p_{\mbox{cl}}(\ket{0})$) or $\ket{1}$ ($p_{\mbox{cl}}(\ket{1})$) as
 \begin{equation}
\label{eq:L}
    V= \frac{p_{\mbox{cl}}(\ket{0})-p_{\mbox{cl}}(\ket{1})}{p_{\mbox{cl}}(\ket{0})+p_{\mbox{cl}}(\ket{1})}
\end{equation}
\noindent and $\alpha = \left[ p_{\mbox{cl}}(\ket{0})+p_{\mbox{cl}}(\ket{1})\right]/2$. 

However, since the probability of detecting a photon in one measurement cycle is very small and $R$ is large, $p (r_n | T_{\chi}, R)$ can be well approximated by a Gaussian distribution \cite{dinani2019, Zohar2023}

\begin{equation}
\label{eq:gauss_approx}
    p (r_i | T_{\chi}, R) \approx \frac{1}{\sqrt{2\pi}\sigma} \exp{\left[ -\frac{\left(r - R \cdot p_D (\tau, T_{\chi}, \beta) \right)^2}{2\sigma^2}\right]}
\end{equation}
with $\sigma = r(R-r)/R$.

Note that as the adaptive update rule in Eq. \eqref{eq:update_rule} does not depend on the outcome but only on the current estimate of $T_{\chi}$ from $p (T_{\chi}|r_{1:n})$, it can be used in the absence of single shot readout.

\begin{figure*}
\includegraphics[height=0.33\textheight]{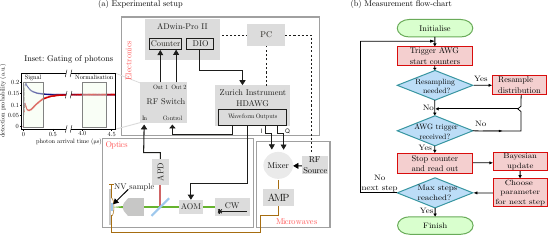}
\caption{\textbf{Experimental setup and measurement flowchart.} (a) Schematic of the hardware (optical, microwave, electronics) in the experimental setup with functional connections. A detailed description can be found in \hyperref[app:exp_setup]{Appendix B}. Acronyms: CW = continuous wave, AOM = acousto-optic modulator, NV = Nitrogen-vacancy centre in diamond, APD = avalanche photodiode, AMP = radio-frequency (RF) amplifier, AWG = Arbitrary Waveform Generator, PC = computer. The computer is used to set the overall measurement and is then not active during the experiment (managed in hard real-time by the AdWin Pro II microcontroller). 
In the inset, we show the histogram of the photon arrival times when the spin is prepared in $\ket{0}$ and $\ket{1}$. The RF switches select the first shaded range, with a difference in photoluminescence intensity between the two spin states, as signal (directed to counter-A) and the second range, with no difference, as background (directed to counter-B) to monitor drifts.   (b) Flow-chart representing the control flow of the adaptive experiments. The control flow of the microcontroller is agnostic to the actual measurement being performed, which is activated by trigger signals. Resampling, if necessary, is performed in parallel to data acquisition (i.e. between the triggering of the AWG with counters starting and the counters stopping) so that no overhead is added to the procedure.}
\label{fig:electronics_flowchart}
\end{figure*}

The experimental setup is sketched in Fig. \ref{fig:electronics_flowchart}(a) (more details in \hyperref[app:exp_setup]{Appendix B}). A real-time microcontroller performs the update of the probability distribution of $T_{\chi}$ via a particle filter and chooses the optimal probing time $\tau$, in a time-scale of about $50 ~\mu$s. The value of $\tau$ is then passed to an arbitrary waveform generator (AWG) which, in real-time, generates the appropriate spin manipulation pulse sequence (including laser and microwave pulses). A computational latency of $50~ \mu$s is a factor $20$ faster than previous real-time quantum sensing experiments at room-temperature \cite{joas2021online}, and much shorter than even the shortest measurement timescales in our experiments (tens of milliseconds, for $\tau \sim 1~\mu$s and $R=10^4$).

\begin{algorithm}[H]

\caption{Adaptive estimation algorithm}\label{alg:adaptive}

    \begin {flushleft}
    \textbf{Input:}\\
    $p_0 (x)$: prior probability distribution for $x=T_{\chi}$;\\
    $K$: number of particles;\\
    $N$: number of epochs;\\
    $a_{LW}$: Liu-West re-sampling parameter;\\
    $t_{RS}$: re-sampling threshold
    \end{flushleft}

\begin{algorithmic}[1]

    \Procedure{AdaptiveEstimation}{$n$, $p_0$, $N$, $a_{LW}$, $t_{RS}$}
    \State draw $\lbrace x_k \rbrace$ from $p_0(x)$
    \State $\lbrace \omega_k \rbrace \gets \lbrace 1/K \rbrace$
    
    \For{$i \in 1.. N$}
        \State $\hat{T}_{\chi} \gets \sum_j \omega_k \cdot x_k$ 
        \State $\tau \gets \xi \cdot \hat{T}_{\chi}$
        \For{$j \in 1.. R$}
            \State $m_j \gets \mbox{EXPERIMENT} (\tau)$
        \EndFor
        \State $r_i \gets \sum_{j=1}^R m_j$

        \State $\lbrace \omega_k \rbrace \gets \lbrace \omega_k \cdot p \left(r_i|T_{\chi}, R \right) \rbrace$
        \State $\lbrace \omega_k \rbrace \gets \lbrace \omega_k/\left(\sum_k \omega_k\right) \rbrace$
        
        \If {$1/ \sum \omega_k^2 < n \cdot t_{RS}$}  
        \State $\lbrace x_k \rbrace \gets \mbox{RESAMPLE} (\lbrace x_k \rbrace, \lbrace \omega_k \rbrace, a_{LW})$
        \State $\lbrace \omega_k \rbrace \gets \lbrace 1/K \rbrace$
    \EndIf
    \EndFor
    \\
    \Return $\hat{T}_{\chi}$

    \EndProcedure
    \\
    \Procedure{RESAMPLE}{$\lbrace x_k \rbrace$, $\lbrace \omega_k \rbrace$, $a_{LW}$}
    \State $\mu \gets \sum_k x_k \cdot \omega_k$
    \State $\sigma^2 \gets \sum_k x_k^2 \cdot \omega_k - \mu^2$
    \State $\mu' \gets a_{LW} \cdot x_k + (1-a_{LW})\cdot \mu$
    \State $\lbrace x_k \rbrace \gets \mbox{NORMAL} (\mu', \sigma^2)$
    \\
    \Return $\lbrace x_k \rbrace$
    \EndProcedure
    \end{algorithmic}
    
    \end{algorithm}

The estimation algorithm is detailed in Algorithm 1. We first choose a discretisation of the distribution of $T_{\chi}$, described by a number $K$ of particles $\lbrace x_k \rbrace$ distributed according to the prior probability $p_0 (T_{\chi})$ (uniform in our case). We initially set the weights of each particle to $1/K$, and $K=100$. At each iteration, we compute the mean $\hat{T}_{\chi}$ of the distribution (line 5) and use it to set the next probing time to $\tau = \xi \cdot \hat{T}_{\chi}$ (line 6). We then perform the selected experiment $R$ times, detecting $r$ photons. The next step updates the probability distribution $p(T_{\chi})$ according to Bayes rule (line 11), normalizing the distribution (line 12).
At this point, if the distribution features large areas with small weights (described by the condition in line 13), is re-sampled according to the Liu-West algorithm \cite{liu2001combined}. We compute the variance  $\sigma^2$ and we then sample new particles (line 25) from a Gaussian distribution with variance $\sigma^2$ and mean:
\begin {equation}
 \mu' = a_{LW} \cdot x_k + (1-a_{LW})\cdot \mu \quad,
\end{equation}
where $a_{LW}$ is the Liu-West parameter, which determines how much the new sampling preserves the original $\lbrace x_k \rbrace$ and how much it reflects the properties (mean $\mu$) of current $p(T_{\chi})$. Resampling is performed in parallel to data acquisition (see Fig. \ref{fig:electronics_flowchart}), so that it does not add any additional delay to the overall measurement time.

In the experiments described in Sec. \ref{sec:exp_results} we compare the adaptive protocols with a non-adaptive alternative where the probing time $\tau$ is chosen randomly within an expected range, known a-priori. The choice of this range affects the performance of the protocols, as adaptive techniques are more effective when there is a large uncertainty in the parameter to be estimated (i.e. for high dynamic range estimation). If the parameter is already known with a good approximation, then settings can already be optimised a-priori, without the need for real-time adaptation. In the experiments presented in Sec. \ref{sec:exp_results} we will select the a-priori parameter range based on typical ranges of variation of decoherence timescales for NV centres in diamond at room temperature.

\subsection{Experimental results}
\label{sec:exp_results}

\begin{figure*}[]
\includegraphics[width=0.8\textwidth]{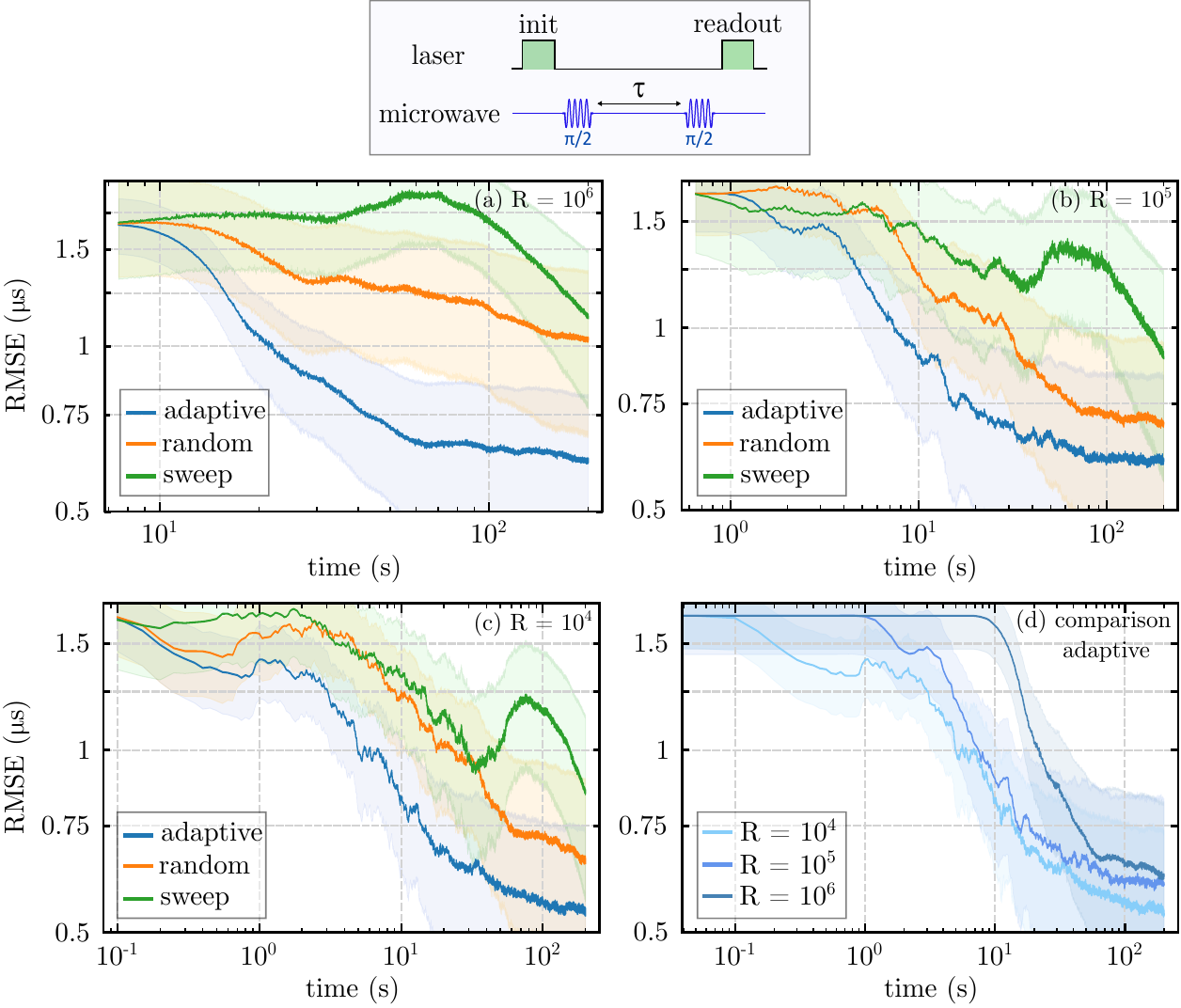}
\caption{ \textbf{Experimental comparison between the adaptive and non-adaptive estimation of the dephasing time $T_2^*$.} The dephasing time $T_2^*$ is measured by a Ramsey experiment. The spin state, initially polarised by a laser pulse in $\ket{0}$, is prepared in an initial superposition state $(\ket{0}+\ket{1})/2$ with a microwave $\pi/2$ pulse. A second $\pi/2$ pulse, after a delay $\tau$, converts the phase information into the population of the $\ket{0}$ state, which is then readout optically. Our adaptive protocol (blue line) outperforms non-adaptive protocols, either sweeping $\tau$ on a pre-determined range (green line), or randomly picking $\tau$ (orange line), for a different number of readout repetitions: $R=10^6$ (a), $R=10^5$ (b) and $R=10^4$ (c). In each sub-plot, we show the root-mean-squared error (RMSE), averaged over 110 experimental protocol repetitions, as a function of the total probing time in seconds.  (d) Comparison of the performance of the adaptive protocol for different values of $R$, with the same experimental data as in (a), (b), and (c) plotted together for ease of comparison. For all plots the parameter range is $T_2^f* \in [1\times 10^{-7}~s, 8\times 10^{-6}~s]$, the ground truth is $2.5~\mu s$ and the shaded areas correspond to $95\%$ confidence intervals.}
\label{fig:2a}
\end{figure*}

Fig. \ref{fig:2a} compares the performance of adaptive and non-adaptive protocols to estimate the static dephasing $T_2^*$ timescale more extensively. We compare three different values of readout repetitions $R$ ($R=10^6$, $R=10^5$, $R=10^4$). Given our experimentally measured values of $p_{cl} (\ket{0}) = 0.0186 \pm 0.0017$, and $p_{cl} (\ket{0}) = 0.0148\pm 0.0016$ photons detected per readout for the two qubit states $\lbrace \ket{0}, \ket{1} \rbrace$, the three different numbers of readout repetitions $R$ correspond, respectively, to mean photon numbers $\langle n \rangle \sim 16700$, $\langle n \rangle \sim 1670$ and $\langle n \rangle \sim 167$. We implement two forms of non-adaptive protocols, one in which $\tau$ is chosen at random within the given range $0-8~\mu$s (based on typical experimentally observed ranges), and one in which $\tau$ is swept uniformly across the range. For a fair comparison, the total number of measurements $N \cdot R$ (with $N$ being the number of epochs) was kept constant for different R, in order to keep to total measurement time fixed.
All curves are averaged over 110 repetitions of the whole estimation sequence in order to obtain the mean performance and the 95\% confidence interval.

\begin{figure}[h!]
\centering\includegraphics[width=0.83\linewidth]{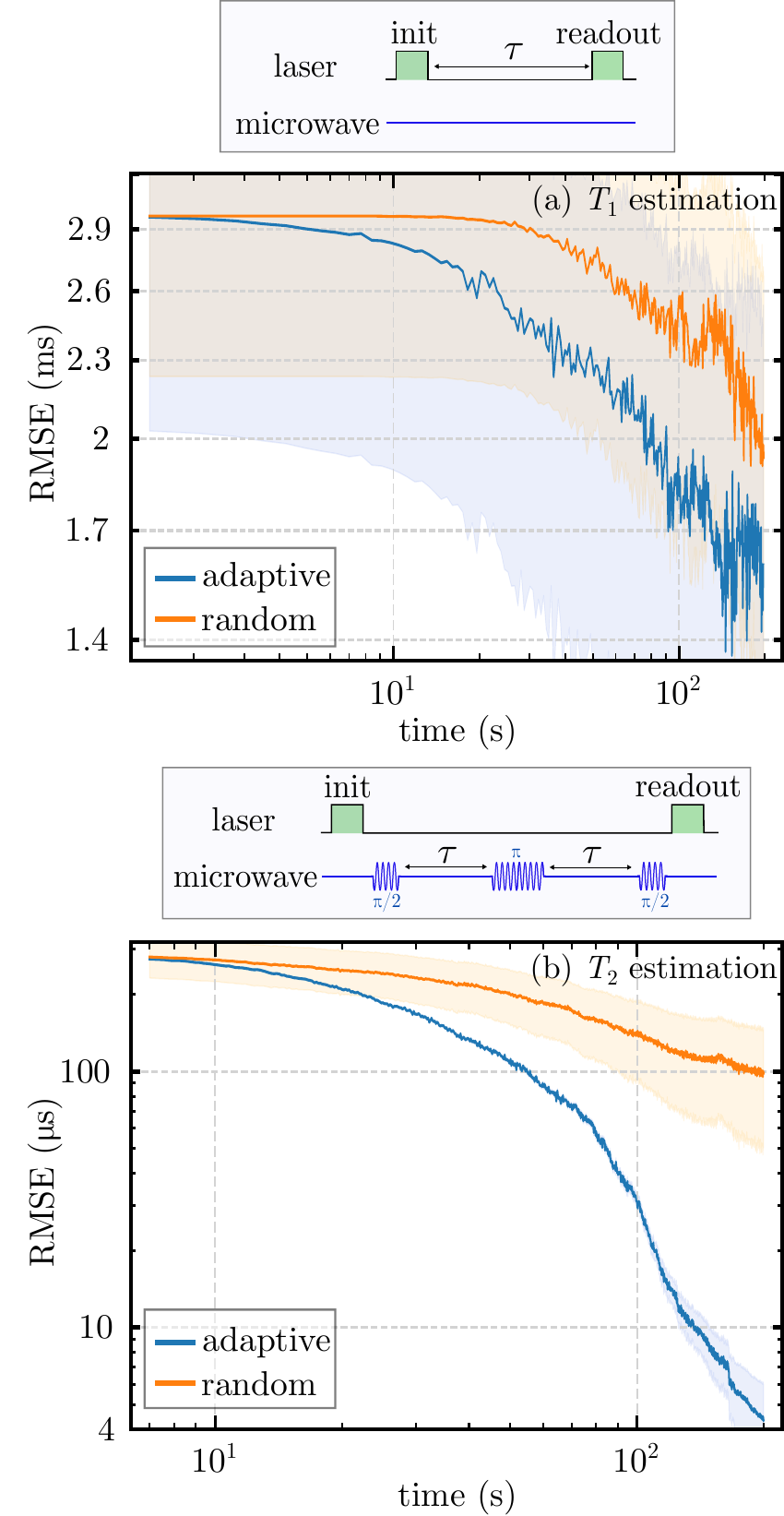} \caption{\textbf{Experimental comparison of adaptive and non-adaptive estimation for $T_1$ and $T_2$.}  In both cases the adaptive choice of $\tau$ is performed maximizing $F$, and the number of readout repetitions is $R=10^4$. (a) Standard deviation as a function of probing time in the estimation of spin relaxation $T_1$ ($\beta = 1$). Each measurement is performed by initializing the electron spin in $\ket{0}$ and detecting the probability of being in $\ket{0}$ after a delay $\tau$. The parameter range is [$1\times 10^{-4}~s, 8\times 10^{-3}~s$] and the ground truth is $1.45~ms$. The standard deviation is averaged over 10 repetitions of the protocol. (b) Standard deviation as a function of probing time in the estimation of the Hahn echo decay time $T_2$ ($\beta = 3/2$). Each measurement is performed by preparing the qubit in $\left( \ket{0}+\ket{1}\right)/2$ and detecting the overlap with the initial state after a time $2 \cdot \tau$, with a spin-flip at time $\tau$ to re-phase the state evolution cancelling the effect of static magnetic uncertainty. The parameter range is [$5\times 10^{-6}~s, 1\times 10^{-3}~s$] and the ground truth is $35~\mu s$. The standard deviation is averaged over 40 repetitions of the protocol. In all plots, shaded regions correspond to the $95\%$ confidence interval. The confidence interval is much wider for $T_1$ estimation as the number of protocol repetitions is quite small (due to the long time required given the ms-scale delays).
}
\label{fig:3}
\end{figure}

Fig. \ref{fig:2a}(a) highlights the comparison of measurement uncertainty versus total measurement time for $R=10^6$ repetitions per experiment, starting from a flat prior corresponding to the same starting point for all three curves. The evolution of the uncertainty is visibly distinct for the three protocols. The adaptive protocol starts learning about the unknown parameter within just a few epochs, leading to a quick drop in the uncertainty compared to the non-adaptive protocols. For the random choice of delay $\tau$ the learning is intermediate in performance between the adaptive protocol and the non-adaptive sweep. The sweep protocol performs particularly badly in the first few estimation epochs, as it performs measurements with probing times much shorter than the decoherence rate, resulting in very little information gained. Once it sweeps across probing times closer to the true decoherence rate, uncertainty is reduced much faster. The protocol with random probing times, on the other hand, samples from the whole expected range uniformly, so that it has a higher probability of retrieving significant information even in the first epochs. The adaptive protocol outperforms both non-adaptive ones, as it quickly learns the optimal probing time to retrieve the most information about the decoherence rate. Comparison between the three experimental curves highlights that, for example, choosing an uncertainty of $1~\mu$s, the adaptive scheme is about $10$ times faster than both the non-adaptive schemes.

Fig. \ref{fig:2a}(b) shows the comparison for $R=10^5$, with a similar trend, and the learning achieved by the adaptive protocol is about $4$ times faster than the random. The sweep protocol performs a factor of $20$ times worse. A similar advantage of the adaptive protocol is also shown for $R=10^4$ [see Fig. \ref{fig:2a}(c)].

In Fig. \ref{fig:2a}(d), a comparison of the performance of the adaptive scheme for different $R$ values, shows that $R=10^4$ (corresponding to $\langle n \rangle \sim167$ photons detected on average) achieves the minimum uncertainty obtained for any measured probing time, highlighting that more frequent adaptive updates are beneficial. Decreasing $R$ even further, in the Gaussian approximation in Eq. (\ref{eq:gauss_approx}), does not improve performance further. Therefore we fix $R=10^4$ for the experiments in the rest of the paper. We also focus only on the version of the non-adaptive protocol with a random choice of $\tau$, as it performs better than parameter sweeping. 

\begin{figure}
    \centering
    \includegraphics[width=1\columnwidth]{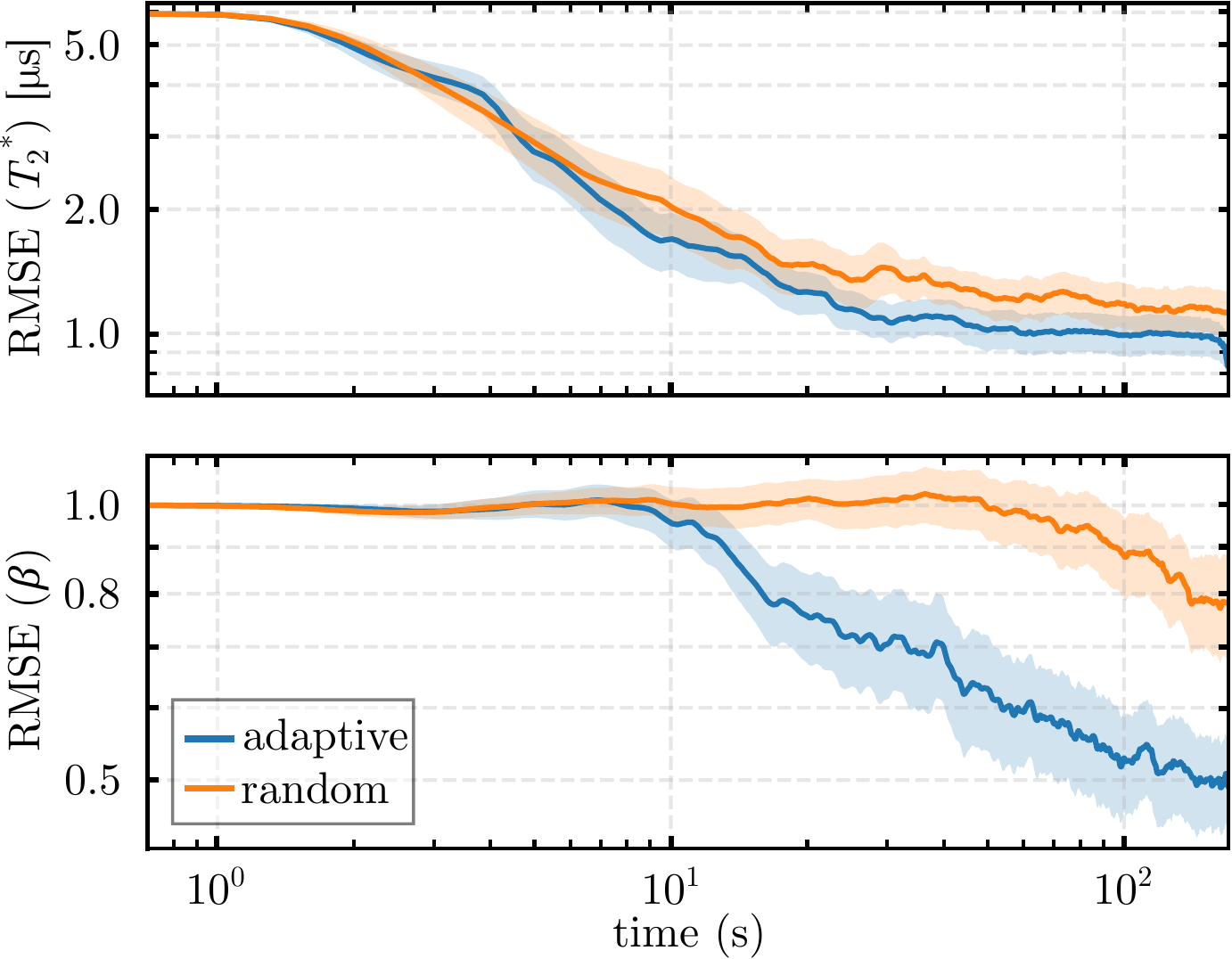}
    \caption{\textbf{Simultaneous experimental estimation of} $\boldsymbol{T_2^*}$ (top) \textbf{and} $\boldsymbol{\beta}$ (bottom). The number of readout repetitions is $R=10^5$. The results are averaged across 140 measurements. The piece-wise linear approximation for the adaptive protocol is used. We set our parameter ranges to be $T_2^* \in [2\times 10^{-7}~s,2\times 10^{-5}~s]$, and $\beta \in [1,5]$. The ground truth values for our system are $T_2^* = 4.15~\mu s$ (a different NV compared to single parameter experiments) and $\beta = 2$. In all plots, the shaded areas correspond to 95\% confidence intervals. }
    \label{fig:2paramexpt}
\end{figure}

Results in Fig. \ref{fig:2a} show that the sweep protocol performs poorly,  due to very little information gain at the start. Therefore in the subsequent measurements (Fig. \ref{fig:3} and Fig. \ref{fig:2paramexpt}) we restricted ourselves to a comparison between the random (non-adaptive) and adaptive schemes. In Fig. \ref{fig:3}, we compare the adaptive and non-adaptive protocols for the estimation of $T_1$ (relaxation) and $T_2$ (echo decay) timescales. In Fig. \ref{fig:3}(top) plot we find that for $T_1$ estimation the adaptive protocol performs  $\sim$ 2 times better than non-adaptive random scheme. This performance is limited by experimental uncertainties, as the long delay time combined with  $R =10^4$ repetitions leads to greater fluctuations over time as opposed to the other estimation sequences $(T_2, T_2^*)$ presented here, where the targeted decoherence timescale in $\mu$s regime. In Fig. \ref{fig:3}(bottom) plot we see that the adaptive protocol outperforms the non-adaptive by significant margin. Such a remarkable gain is achieved as our knowledge about $T_2$ for an NV centre in buld diamond is typically more uncertain apriori than the value of $T_1$. In both cases, the adaptive strategy greatly outperforms the non-adaptive protocol.

We verify the performance of multi-parameter estimation [see Eq. (\ref{eq:2paramapprox})]  using Ramsey measurement ($T_{\chi}=T_2^*$). The choice of optimal sensing time $\tau$ is based on using the piecewise linear approximation, discussed in \hyperref[app:exp_setup]{Appendix B}. In Fig. \ref{fig:2paramexpt}, we compare the adaptive protocol against random measurement scheme, plotting the estimation error of both $T_2^*$ and $\beta$ versus total probing time, demonstrating a clear advantage for the adaptive algorithm. The gain observed in the $T_2^*$ plot is $\sim$ 10 times for adaptive estimation over the random non-adaptive scheme. While $\beta$ estimation plot shows a gain of $\sim$ 20 times for the adaptive scheme. For this experiment the number of particles (K) over which the Bayesian update is performed was increased to 500 to better sample the two parameter space.

\begin{figure}[h!]
\includegraphics[width=1\columnwidth]{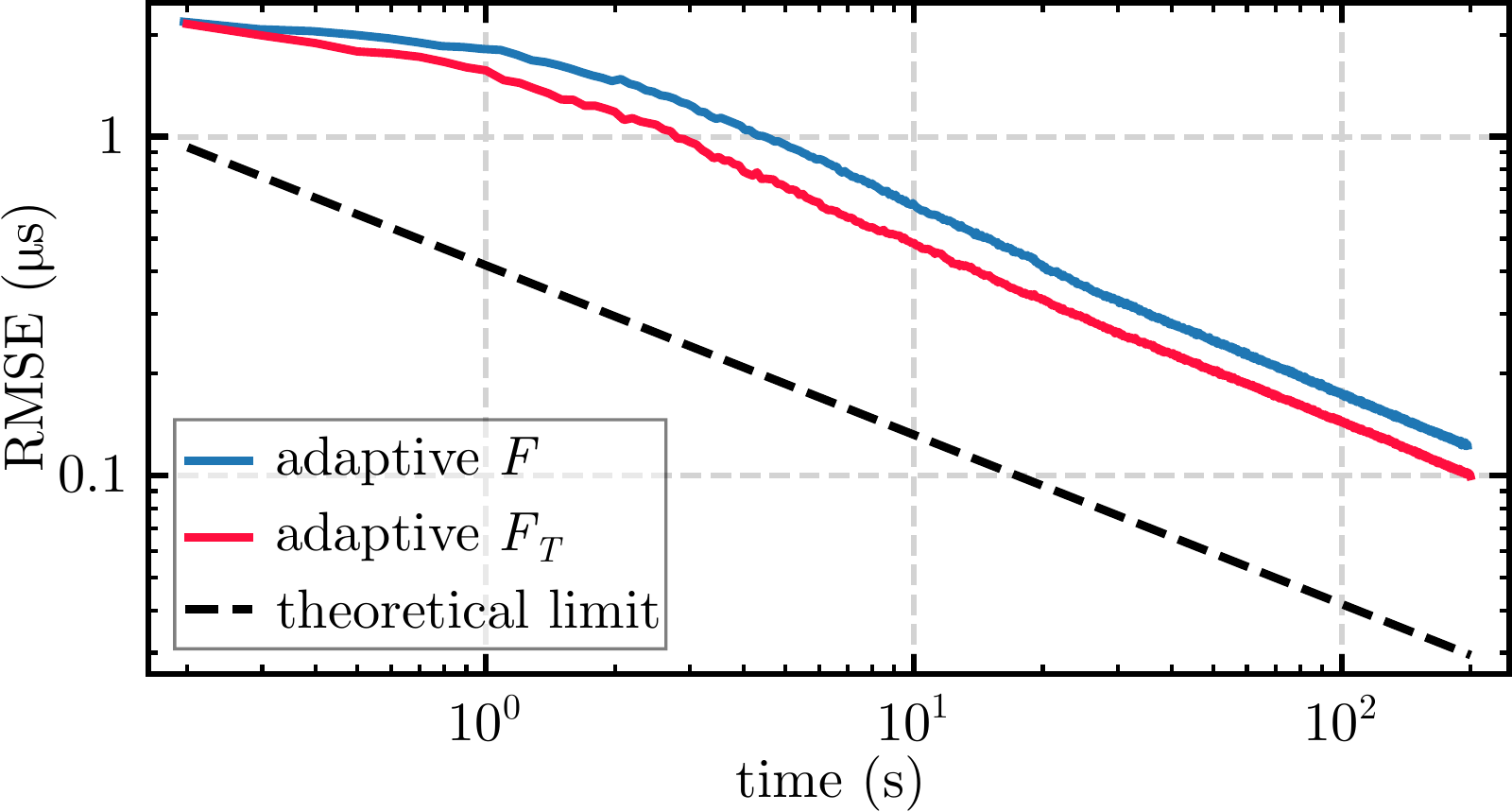}
\caption{ {\bf Experimental comparison between maximizing $F$ and $F_T$.} An adaptive choice computed maximizing the Fisher information rescaled by the total probing time [see Eq. (\ref{eq:F_T})] achieves a smaller uncertainty for a given probing time than simply maximizing the Fisher information. The experimental estimations of $T_2^*$ ($\beta = 2$) are performed with $R = 10^4 $, with parameter range $T_2^* \in [1\times 10^{-7}~s, 8\times 10^{-6}~s]$, ground truth $2.5~\mu s$ and averaged over 35 repetitions of the whole protocol. The dashed black line corresponds to the theoretical limit given by $F_E$ in Eq. (\ref{eq:F_E}). Both experimental curves match the scaling dictated by the CRLB for $F_E$.}
\label{fig:4}
\end{figure}

Fig. \ref{fig:4} compares adaptive estimation of the dephasing time $T^*_2$ when maximizing the BIM  (related to $F$) or the BIM rescaled by the probing time (related to $F_T$). As discussed, this results in different multiplication factor $\xi$ for the adaptive choice of $\tau$ rule as given in Table \ref{FI_table}. Fig. \ref{fig:4} shows that maximisation of $F_T$ is a factor $\sim$2 faster than maximisation of $F$, as a function of the total probing time, as expected from Sec. \ref{sec:opt_sens}. It is worth noting that both protocols shown in Fig. \ref{fig:4} eventually reach the same scaling as the Fisher Information, but with different offsets. This experimental result is consistent with the theoretical prediction from Fig. \ref{fig:T2star_sims_SSRO}. Similarly, the experimental results shown earlier display strong correspondence with theoretical predictions, providing a robust base for the improved performance of adaptive protocols.

Generally, while all experimental datasets reproduce the predicted trends, they do not appear to reach the theoretical limits. We believe this is due to residual experimental imperfections or fluctuations not accounted for by the likelihood function we use in the Bayesian inference process, which could be incorporated through data-driven or ``gray-box" approaches \cite{youssry2022experimental}.

\section{Discussion and outlook}
Using an adaptive Bayesian approach based on high-speed electronics, we have compared approaches to estimating decoherence timescales and decay exponents, showing that real-time adaptation of the probing time $\tau$ significantly outperforms non-adaptive approaches. We have also investigated different adaptation heuristics, demonstrating an advantage in sensitivity optimisation in situations where time is the resource to be minimised.

Real-time adaptation of experimental settings is most valuable when there is a large uncertainty in the parameter to be estimated. In this case, pre-optimising the experiment offline is not trivial, as the optimal settings typically depend on the unknown value of the parameter we are measuring. On the other hand, real-time adaptation is less important when the parameter to be estimated varies on a very small range, as in this case the experimental settings can be optimised offline based on the strong available prior information. In this paper, we have chosen reasonable ranges for the decoherence timescales and exponents we estimate, based on our own experience in the field and the variability observed in the literature.

The techniques demonstrated here will directly improve the performance of quantum relaxometry, where measuring the decoherence of a qubit is used to extract useful information from noise in the environment. An example is the use of spins in nano-diamonds to measure the concentration of radicals, exploiting the fact that they induce magnetic noise that shortens the sensor spin $T_1$ timescale. This has been applied, for example, to probe specific chemical reactions \cite{rendler2017optical} or the concentration of radical oxygen species inside living cells \cite{nie2021quantum,nie2021quantum_}. The time required to take a measurement limits the bandwidth over which fast signal variations can be tracked, a very important parameter when dealing for example with biological processes in living cells.

Our technique can be readily extended to more complex experiments and pulse sequences, for example, double electron-electron resonance (DEER) or electron-nuclear double resonance (ENDOR) where measurements on electronic spins are used to infer the dynamics and decoherence of other electronic or nuclear spin. DEER experiments on single NV centres close to the diamond surface, for example, reveal changes in decay exponent that elucidate the physics of dark spins on the diamond surface \cite{nathalie_PRXQ}.

While our experiments are performed using a single electronic spin associated with an NV centre in diamond, our approach and our analysis are very general and can be readily applied to any qubit system and to different applications. For example, our approach could be used for fast characterisation of quantum memories, or multi-qubit quantum processors.

In summary, we have shown that real-time adaptive estimation of decoherence, through a simple analytical optimisation heuristic, leads to a considerable speed-up compared to non-adaptive schemes. Extending this approach to more complex settings, such as towards the detection of multiple individual nuclear spins and nanoscale magnetic resonance \cite{zopes2018three,abobeih2019atomic,jung2021deep,budakian2023roadmap} , holds the promise to deliver even greater advantages.

\section*{\label{app:derivations} Appendix A: derivation of adaptive rules}
As discussed in Sec. \ref{sec:single_par_estimation_theory}, we optimize the settings for the next measurement by considering an approximation of the Bayesian information matrix (BIM) (a $1\times 1$ matrix, in the case of a single parameter) \cite{van2007bayesian} which links to the classical Fisher information \cite{steven1993fundamentals}.

For a binary outcome $m = 0$ or $1$, the classical Fisher information for the decoherence timescale $T_{\chi}$ is
\begin{equation}
    F = \sum_{m = 0, 1} \left[ \frac{\partial}{\partial T_{\chi}} \log p \left(m|T_{\chi} \right) \right]^2 p \left(m|T_{\chi} \right) \quad,
\end{equation}
where $p(m|T_{\chi})$ is the probability to detect outcome $m$ given a decoherence timescale $T_{\chi}$ described by Eq. (\ref{eq:likelihood}).

The classical Fisher information $F$ is then given by:
\begin{equation}
\label{eq:FI}
    F(\tau) = \frac{\beta^2 (\tau/T_{\chi})^{2\beta}}{T_{\chi}^2 \left[ e^{\left(2\tau/T_{\chi}\right)^{\beta}}-1\right]} \qquad .
\end{equation}

The Fisher information $F$ is inversely proportional to the Cram\'er-Rao lower bound (CRLB) of $T_{\chi}$, which represents the minimum reachable variance for any (unbiased) estimator of $T_{\chi}$. Thus, maximizing $F$ with respect to $\tau$ is expected to improve our estimate of $T_{\chi}$. The value of $\tau$ that maximizes Eq. (\ref{eq:FI}) {\it i)} needs to be approximated numerically and more importantly {\it ii)} depends on $T_{\chi}$ which is unknown [this dependency is not explicit in the notation $F(\tau)$ but visible in Eq. (\ref{eq:FI})]. In this context, where prior information is known about $T_{\chi}$ from previous measurements [see Eq. (\ref{eq:probability})], the Bayesian CRLB \cite{van2007bayesian} is a more adapted criterion as it averages the Fisher information over the high-density region of the prior distribution of $T_{\chi}$. In the single-parameter setting considered here, the BIM is given (see \cite{van2007bayesian}) by 
\begin{eqnarray}
\label{eq:BIM}
F_{\mathcal{B}}(\tau)=\textrm{E}_{T_\chi}\left[F(\tau) + \dfrac{\partial^2 \log(P(T_\chi|m_{1:(n-1)}))}{\partial^2 T_\chi}\right],
\end{eqnarray}
where $\textrm{E}_{T_\chi}\left[\cdot\right]$ denotes the statistical expectation with respect to $P(T_\chi|m_{1:(n-1)})$. The Bayesian CRLB (BCRLB) is the inverse of $F_{\mathcal{B}}(\tau)$ and bounds the performance [mean squared error (MSE)] of estimators of $T_{\chi}$ \cite{van2007bayesian}. Instead of maximizing the information gain or minimizing the MSE, we use the BIM as a proxy reward function to be maximised when adapting $\tau$. First, it is worth noting that the second term in the expectation in Eq. (\ref{eq:BIM}) does not depend on $\tau$. Thus, it it sufficient to maximize $\textrm{E}_{T_\chi}\left[F(\tau)\right]$. To overcome this intractable integral, we use 
\begin{eqnarray}
\label{eq:F_B_appendix}
    F_{\mathcal{B}}(\tau) \approx \frac{\beta^2 (\tau/\hat{T}_{\chi})^{2\beta}}{\hat{T}_{\chi}^2 \left[ e^{\left(2\tau/\hat{T}_{\chi}\right)^{\beta}}-1\right]}=\hat{F}_{\mathcal{B}}(\tau)
\end{eqnarray}
where $\hat{T}_{\chi}$ is a point estimate of $T_{\chi}$. This corresponds to replacing, in the integral $\textrm{E}_{T_\chi}\left[F(\tau)\right]$, the distribution $P(T_\chi|m_{1:(n-1)})$ by a Dirac Delta function centered at $\hat{T}_{\chi}$. If the prior distribution $P(T_\chi|m_{1:(n-1)})$ is concentrated around $\hat{T}_{\chi}$, one option is to choose $\hat{T}_{\chi}$ as the most likely value of $T_{\chi}$ \cite{zhang2018improving}. Here we used instead the prior mean as it can be more robust when $P(T_\chi|m_{1:(n-1)})$ has a large variance and/or is skewed. Note that here we do not specify how $P(T_\chi|m_{1:(n-1)})$ and its moments are computed. This is detailed in Sec. \ref{subsec:Algo}.

As discussed in Sec. \ref{sec:single_par_estimation_theory}, we find an approximate maximum of Eq. (\ref{eq:F_B_appendix}) as
\begin {equation}
\label {eq:update_rule}
 \tau_{opt} \approx \xi \cdot \hat{T}_{\chi}
\end{equation}, where $\xi$ is a parameter that depends on $\beta$.  

\section*{\label{app:exp_setup} Appendix B: experimental setup}

{\bf Sample.} The sample consists of an electronic-grade CVD diamond plate (Element Six). Isolated NV centres with good spin coherence are created by the laser-writing of vacancies at an array of sites in the diamond followed by thermal annealing to form NV centres with background nitrogen impurities.

{\bf Optics.} The system is based on a custom-made confocal microscope. The sample, mounted on a custom PCB on a Newport stage, is imaged through an oil immersion objective (Olympus RMS100X-PFO). Photoluminescence is excited by a CW 532nm diode-pumped solid state laser (CNI MLL-U-532-200mW), pulsed by an acousto-optic modulator (Isomet 1250C-829A) in double-pass configuration. The laser beam is scanned across the sample by a galvo mirror pair (Thorlabs GVS012/M), using a 4f optical system. The photoluminescence is collected through a dichroic mirror (Semrock LP02-633RU-25) and detected by a fibre-coupled single photon avalanche photodiode (Excelitas SPCM-AQRH-15-FC). The optical setup is mounted inside a styrofoam box and temperature stabilised by a Peltier element controlled by a PID loop (Meerstetter TEC-1092), down to 10 mK rms.

{\bf Magnetic field.} A magnetic field of 460 gauss, aligned with the NV axis, is applied by a permanent SmCo magnet mounted on a 3D motorised translation stage (Standa 8MT173-30DCE2-XYZ). The magnetic field is selected at the NV centre excited-state level anti-crossing to polarize the $^{14}$N nuclear spin, so that no hyperfine lines are present in the Ramsey experiments. The magnet is encased in a 3D-printed holder, and thermally stabilised by a Peltier element controlled by a PID loop (Meerstetter TEC-1091), down to 10 mK rms.

{\bf Microwaves.} Microwaves are generated by single-sideband modulation of a tone at 1.53 GHz, from a vector source (RohdeSchwarz SMBV100A), with a low-frequency signal (40 MHz) from a Zurich Instruments HDAWG4 arbitrary waveform generator (2.4 GSa/s, 16 bits vertical resolution, 750 MHz signal bandwidth). The modulated microwave signal is amplified (Amplifier Research 15S1G6, 0.7-6 GHz, 15 W) and directed to a wire across the sample, through a circulator (Aaren Technology 11B-GX017027G-AF), to create an RF magnetic field to drive the ($m_s = 0 \Longleftrightarrow m_s = -1$) electron spin resonance. 

{\bf Adaptive electronics.} The real-time experiment optimisation is handled by a hard real-time micro-controller (AdWin Pro II, J\"{a}ger Computergesteuerte Messtechnik). Electrical pulses from the photon detectors are sent to a pair of switches (Minicircuit ZASW-2-50DRA+), that re-direct the window A ($\left[t_0, t_0 + 270 ~\mathrm{ns} \right]$) to counter-1 in the AdWin and the window B ($\left[t_0 + 4 ~\mathrm{\mu s}, t_0 + 4 ~\mathrm{\mu s} + 400 \mathrm{ns} \right]$) to counter-2 ($t_0$ is the start time of the excitation laser pulse). Counts in window-A correspond to photon counts that discriminate the spin signal [see Fig. \ref{fig:electronics_flowchart}(a) inset], since $m_s=0$ and $m_s = \pm 1$ exhibit different photo-luminescence intensities. Counts in window-B are not spin-dependent and are used to monitor the system's stability.

The microcontroller reads the values in the two counters and uses the signal counts in Window-A to perform a Bayesian update and to select the optimal probing time $\tau$ for the next measurement. The value f $\tau$ is passed, as an 8-bit integer to the HDAWG through its digital IO port. For each batch of $R$ measurements, the HDAWG reads the value from the DIO port and constructs the next pulse sequence accordingly. The HDAWG generates control pulses for the acousto-optic modulator, single-sideband modulation of the microwave signal and to control the switch to re-direct electrical pulses from the detector to the microcontroller counters.

{\bf Details of micro-controller (MCU) operation flow.} The operation of the MCU is described by the flow-chart in Fig. \ref{fig:electronics_flowchart}(b).  At each iteration, the MCU triggers the AWG, which delivers all the control pulse sequences, and starts the counters. If resampling is needed, it is started at this point and carried out in parallel to data acquisition so it does not add any additional overhead. Once the MCU receives a trigger back from the AWG, signalling that all pulse sequences have been executed, it stops and reads the counters, which have accumulated the total count rate for all $R$ repetitions. The MCU then performs the Bayesian update of $p(T_{\chi})$ and selects the next $\tau$, sending the value to the AWG to compile the next qubit control sequence.

The timing of the experiment was benchmarked with respect to the internal clock of the microcontroller, finding $T_{\mu \mbox{s}} \sim 0.255 \cdot n$, where $n$ is the number of particles discretising $T_{\chi}$, i.e. about $50~\mu$s for $n=200$ particles.

\begin{figure*}
\centering
\includegraphics[width=1.0\textwidth]{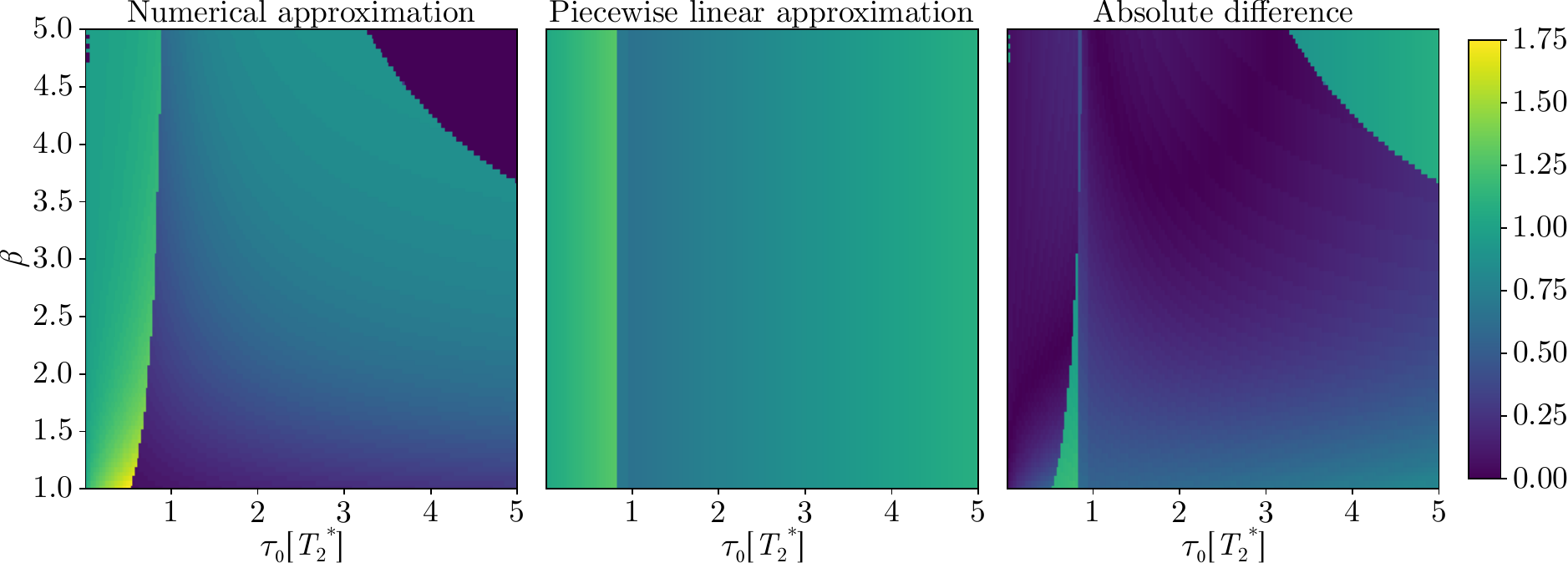}
\caption{\textbf{Piece-wise linear approximation accuracy.} Plots demonstrate the optimal choice of next measurement time ($\tau_1$) vs the current estimate of coherence decay exponent $\beta$ and previous measurement time $\tau_0$. Both measurement times are normalised to the current estimate of $T_2^*$. The accuracy of the numerical approximation (left) is limited by the 64-bit resolution of the floating point used in the calculation, resulting in a region in the upper right of the numerical approximation plot that was manually set to zero. The piece-wise linear approximation given by Eq. (\ref{eq:2paramapprox}) is plotted in the centre, and the absolute difference between these two approximations is given in the right-hand side plot.} 
\label{fig:piecewise_approx_compare}
\end{figure*}

{\bf Details of multiparameter approximate adaptive rule.} 
As we could not find an analytical solution to create an update rule, instead we evaluate the equation numerically across our region of interest and fit an approximation. The results are shown in Fig. \ref{fig:piecewise_approx_compare}. We numerically evaluate Eq. (\ref{eq:detBIM2d4o}), modified to include terms to quantify measurement readout occurring over R repetitions in the absence of single shot readout. We aim to find the simplest approximation to replicate its result. Observation of the steep jump in value just below $\tau_0=1$ suggested the use of a piecewise polynomial fit in $\tau_0$ and $\beta$. Using the polyfit function of the numpy package we find optimal parameters for linear fits in two separate regions. In between these regions, a heuristic choice is made to pick a measurement time $0.7\times T_2^*$. Simulation of experiments using this piecewise linear approximation rule showed similar performance to the use of the full numerical approximation. 

\section*{acknowledgments}

We express our gratitude to Mete Atatüre (University of Cambridge) for valuable discussions and appreciate the technical support provided by Andrea Corna and Jelena Trbovic (Zurich Instruments) as well as Heinz Beimert (Jäger Messtechniek GmbH). This work is funded by the Engineering and Physical Sciences Research Council (EP/S000550/1, EP/V053779/1 and through the UK Quantum Technology Hub in Quantum Imaging EP/T00097X/1), the Leverhulme Trust (RPG-2019-388) and the European Commission (QuanTELCO, grant agreement No 862721). We also acknowledge the support provided by a Rank Prize `Return to Research' grant. C. B. and A. F. are jointly supported by the `Making Connections' Weizmann-UK program. C. Bekker is supported by a Royal Academy of Engineering Research Fellowship (No. RF2122-21-129). G. W. M. is supported by the Royal Society (RGF$\backslash$EA$\backslash$180311 and UF160400), by the UK EPSRC (EP/V056778/1) and by the UK STFC (ST/W006561/1 and ST/S002227/1). G. W. M and J. S. are jointly supported by the EPSRC grant EP/T001062/1.

\end{document}